\pdfoutput=1
\documentclass[a4paper,american,floatfix,pdftex,superscriptaddress,twoside,%
aip,sd,%
citeautoscript,
longbibliography,%
reprint,twocolumn
]{revtex4-1}%
\usepackage{amsfonts,amsmath,amssymb}
\usepackage[T1]{fontenc}
\usepackage{graphicx}%
\usepackage[utf8]{inputenc}
\usepackage{microtype}
\usepackage{titlesec}
\usepackage[obeyFinal,textsize=scriptsize]{todonotes}
\usepackage{xspace}
\usepackage{hyperref, hypernat}
\usepackage[todos]{CMImacros}

\graphicspath{{./figure/}} 
\hypersetup{citebordercolor=yellow,linkbordercolor=red,urlbordercolor=blue}

\titleformat{\paragraph}[runin]{\bfseries\vspace{1em}}{}{1em}{\hspace{-\parindent}}

\makeatletter%
\renewcommand*{\@fnsymbol}[1]{\ensuremath{\ifcase#1\or \|\or *\or **\or \mathparagraph\or
      \mathsection\or \dagger\or \ddagger\or \dagger\dagger \or \ddagger\ddagger \else\@ctrerr\fi}}
\makeatother

\newcommand*{\edfref}[1]{\hyperref[#1]{Extended Data Figure~\ref{#1}}\xspace}
\newcommand*{\sifref}[1]{Fig. S#1\xspace}
\newcommand*{\sifrefs}[1]{Figs. S#1\xspace}

\newcommand{\cfeldesy}{\affiliation{Center for Free-Electron Laser Science, Deutsches
      Elektronen-Synchrotron DESY, Notkestraße~85, 22607 Hamburg, Germany}}%
\newcommand{\uhhcui}{\affiliation{Center for Ultrafast Imaging, Universität Hamburg, Luruper
      Chaussee~149, 22761 Hamburg, Germany}}%
\newcommand{\uhhphys}{\affiliation{Department of Physics, Universität Hamburg, Luruper Chaussee~149,
      22761 Hamburg, Germany}}%
\newcommand{\ruimm}{\altaffiliation[Present address: ]{Radboud University, Institute for Molecules
      and Materials, Heijendaalseweg 135, 6525 AJ Nijmegen, The Netherlands}}%
\newcommand{\jkemail}{\email[Corresponding author. Email:~]{jochen.kuepper@cfel.de}}%
\newcommand{\cmiweb}{\homepage[\mbox{website}:~]{https://www.controlled-molecule-imaging.org}}%

\begin{document}
\title{Controlled beams of shockfrozen, isolated, biological and artificial nanoparticles}%
\author{Amit K.\ Samanta}\cfeldesy%
\author{Muhamed~Amin}\cfeldesy%
\author{Armando~ D.\ Estillore}\cfeldesy%
\author{Nils~Roth}\cfeldesy\uhhphys%
\author{Lena~Worbs}\cfeldesy\uhhphys%
\author{Daniel A.\ Horke}\ruimm\cfeldesy\uhhcui%
\author{Jochen Küpper}\jkemail\cmiweb\cfeldesy\uhhphys\uhhcui%
\date{\today}%
\begin{abstract}\noindent
   X-ray free-electron lasers (XFELs) promise the diffractive imaging of single molecules and
   nanoparticles with atomic spatial resolution. This relies on the averaging of millions of
   diffraction patterns of identical particles, which should ideally be isolated in the gas phase
   and preserved in their native structure. Here, we demonstrated that polystyrene nanospheres and
   Cydia pomonella granulovirus can be transferred into the gas phase, isolated, and very quickly
   shockfrozen, \ie, cooled to 4~K within microseconds in a helium-buffer-gas cell, much faster than
   state-of-the-art approaches. Nanoparticle beams emerging from the cell were characterized using
   particle-localization microscopy with light-sheet illumination, which allowed for the full
   reconstruction of the particle beams, focused to $<100~\um$, as well as for the determination of
   particle flux and number density. The experimental results were quantitatively reproduced and
   rationalized through particle-trajectory simulations. We propose an optimized setup with cooling
   rates for few-nanometers particles on nanoseconds timescales. The produced beams of shockfrozen
   isolated nanoparticles provide a breakthrough in sample delivery, \eg, for diffractive imaging
   and microscopy or low-temperature nanoscience.
\end{abstract}
\maketitle

\section{Introduction}
\label{sec:introduction}
Nanometer objects are of extraordinary importance in nature, for example in the complex biological
machinery of viruses~\cite{Raoult:NatRevMicBio6:315}. Furthermore, the 21st century has been hailed
as the ``age of nanotechnology'', with the advent of, \eg, novel nanomaterials, such as quantum-dot
light emitting diodes~\cite{Shen:NatPhoton13:192} and nanomedicine~\cite{Bao:AnnuRevBioEng15:253}.
Understanding the fundamental functionality of these systems requires high-resolution structural
information. Recent years have seen phenomenal progress in this area. One pioneering approach to
measure direct structural information from isolated nanoparticles is single-particle diffractive
imaging (SPI), enabled by the advent of x-ray free-electron lasers
(XFELs)~\cite{Neutze:Nature406:752, Bogan:NanoLett8:310, Barty:ARPC64:415}. This promises the
recording of atomically-resolved structures from isolated nanoobjects without the need for large,
highly-ordered crystalline samples~\cite{Bogan:NanoLett8:310, Neutze:Nature406:752}. It relies on
recording a series of two-dimensional diffraction images from randomly oriented isolated particles,
which can then be assembled \emph{in silico} to a three dimensional (3D) diffraction volume and the
structure reconstructed. Since the first demonstration of this approach a decade
ago~\cite{Chapman:NatMater8:299}, several significant steps in experimental
procedures~\cite{Seibert:Nature470:78, Ekeberg:PRL114:098102} and data
analysis~\cite{Ayyer:Nature530:202} have pushed the achievable resolution to below
10~nm~\cite{Hosseinizadeh:NatMeth14:877}.

A further technique for direct structural imaging of nanometer-sized objects is cryo-electron
microscopy (CEM), where several recent breakthroughs have enabled single-particle structure
determination to sub-nanometer resolution~\cite{Fernandez:Nature537:339, Sugita:Nature563:137}.
Unlike SPI, CEM images a single nanoparticle, immobilized and shock-frozen onto a support. This
sample preparation using the plunge-freezing approach is a crucial step of CEM
success~\cite{Drulyte:ActaCryst74:560}. However, various issues with the technique have been
discussed~\cite{Drulyte:ActaCryst74:560}.

In contrast to CEM, the SPI approach images isolated particles \emph{in vacuo}, \ie, without any
mechanical sample support. However, due to its diffraction-before-destruction
approach~\cite{Neutze:Nature406:752}, it requires the imaging of millions or billions of identical
particles to allow the reconstruction of the 3D structure. One of the major challenges for improving
the achievable resolution is the reproducibility of the target, \ie, the stream of isolated single
particles probed by the XFEL. To date, these experiments have been conducted with room temperature
aerosols, in an attempt to keep the biological systems studied under native-like conditions. This
approach also leads to a dynamical exploration of the conformational landscape. It demands the
collection of a very large dataset for SPI experiments, which then have to be analyzed for
structures in terms of conformations and spatial orientations. Eventually, this will also limit the
achievable resolution for a given measurement time. Moreover, these experiments are often struggling
with limited hit rates and limited availability of time for experiments at XFEL facilities, which
represent a major obstacle for the collection of a sufficiently large dataset required for a
high-resolution reconstruction.

\begin{figure*}
   \includegraphics[width=\linewidth]{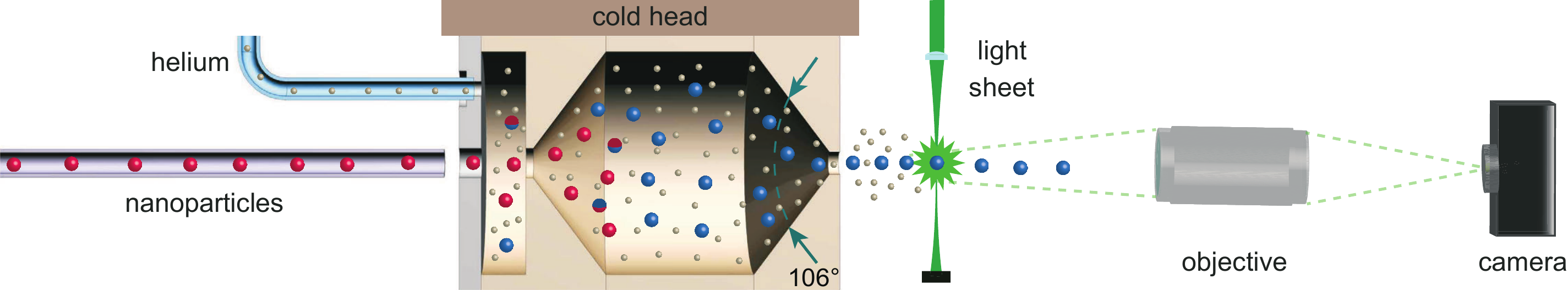}
   \caption{\textbf{Schematic of the experimental setup}. Aerosolized nanoparticles (red spheres)
      are transported into the cryogenically cooled buffer-gas cell, where they collisionally
      thermalize (blue spheres) with precooled helium (brown spheres). They exit the cell forming a
      beam of cold nanoparticles, which is characterized using a single-particle localization
      microscope. The buffer-gas cell has detachable conical entrance and exit endcaps with a full
      opening angle of \degree{106}~[\onlinecite{Singh:PRA97:032704}].}
   \label{fig:schematic}
\end{figure*}

Here, we propose and demonstrate a novel sample preparation using rapidly shock-frozen beams of,
potentially hydrated, isolated nanoparticles. Using a cryogenic buffer-gas cell, nanoparticles were
rapidly cooled on a microsecond timescale, sufficiently fast to prevent denaturation, and then
extracted into a collimated particle stream in vacuum. The produced high-density beams make an ideal
target for SPI experiments and, furthermore, are highly amenable to further control, \eg, by
external electric or acoustic fields. The cryogenic-temperature samples will allow one to spatially
separate conformers~\cite{Filsinger:PRL100:133003, Chang:IRPC34:557, Teschmit:ACIE57:13775,
   Helden:Science267:1483, Lanucara:NatChem:6:281}, to strongly align and orient the particles in
the laboratory frame~\cite{Spence:PRL92:198102, Holmegaard:PRL102:023001}, or to produce very high
densities through the focusing of the particles with external
fields~\cite{Eckerskorn:PRAppl4:064001, Li:PRAppl11:064036}.

Our development provides touch-free shock-frozen sub-10~K focused particle beams of artificial and
biological nanoparticles. Particles were aerosolized from solution at room temperature using a
gas-dynamic virtual nozzle~\cite{DePonte:JPD41:195505, Beyerlein:RSI86:125104} and transported into
a cryogenically-cooled helium-filled buffer-gas cell, in which isolated nanoparticles were quickly
cooled through collisions with the cold helium gas. Buffer-gas cooling is an established technique
in atomic and molecular physics~\cite{Hutzler:CR112:4803}, but had so far not been applied to
systems with more than a few tens of atoms~\cite{Piskorski:CPC15:3800, Kamrath:ACR51:1487}. We
demonstrate its applicability to shock-freeze polystyrene spheres (PS) of 220~nm and 490~nm diameter
as well as the native occlusion bodies (OBs) of \emph{Cydia pomonella} granulovirus (CpGV) particles
with a size of approximately $265\times265\times445~\text{nm}^3$~\cite{Gati:PNAS114:9}. The
shock-frozen particles were extracted from the buffer-gas cell and formed a collimated or focused
nanoparticle beam. Here, individual particles were detected using single-particle-localization
microscopy~\cite{Awel:OptExp24:6507}. Measured particle distributions for different helium-flow
conditions were well reproduced by particle-trajectory simulations, which furthermore allowed us to
extract cooling rates and times.

\section{Methods}
\label{sec:methods}
\subsection{Experimental details}
A schematic of our experimental setup is shown in \autoref{fig:schematic}. It consisted of four main
parts: an aerosolization chamber, a differentially
pumped transport tube, the cryogenically-cooled buffer-gas cell, and a detection region. Isolated
nanoparticles were created by aerosolising aqueous solutions using a gas-dynamic virtual
nozzle~\cite{DePonte:JPD41:195505, Beyerlein:RSI86:125104}. We have used polystyrene spheres of
220~nm (Alfa Aesar, $220\pm17.3$~nm) and $490$~nm (Molecular Probes, $490\pm15$~nm) with a
concentration of $5\times10^{10}$~particles/ml. CpGV samples were produced following a known
protocol~\cite{Gati:PNAS114:9}. Leaving the aerosolization chamber, the particles passed through a
set of two skimmers ($\varnothing_1=0.3$~mm, $\varnothing_2=0.5$~mm) placed $2$~mm apart. The region
between the skimmers was evacuated to remove background gases from the aerosolization processes,
\eg, helium and water, to avoid ice formation and clogging of the BGC inlet and outlet. The
particle stream then entered a transport tube, with a typical pressure of 10~mbar during operation.
The warm isolated nanoparticles were introduced into the buffer-gas cell using a 10~cm long
stainless steel capillary with an inside diameter of 800~\um. The complete aerosol generation and
transport assembly is attached to the vacuum chamber using a three-dimensional position manipulator,
allowing precise alignment of the capillary to the 2~mm buffer-gas cell inlet. During the experiment
the capillary tip was located 7~mm outside the cell entrance aperture. The buffer-gas cell was
located in the main vacuum chamber, maintained at a pressure below $10^{-6}$~mbar by a
turbomolecular pump (Pfeiffer Vacuum HiPace 2300). It was attached to a 2-stage pulse-tube
refrigerator (Sumitomo RP082E2) with typical operating temperatures of 29~K and 3.6~K, shielded from
thermal radiation by aluminium and copper heat shields attached to the cooling stages. Coconut
charcoal attached to the second stage radiation shield provides additional pumping capacity. The
buffer-gas cell itself was a hollow copper cylinder ($\varnothing=3$~cm, $2$~cm length) with
detachable copper endcaps for both entrance and exit side. We used conical endcaps with an opening
angle of \degree{106}~\cite{Singh:PRA97:032704}. Inside the buffer-gas cell, the room-temperature
nanoparticles underwent rapid collisional thermalization with the 4~K cold helium gas at typical
densities of $\ordsim10^{16}~\text{cm}^{-3}$. The cooled nanoparticles were extracted through an
exit aperture of 2~mm diameter into high vacuum, $p<10^{-6}$~mbar, forming a collimated/focused
particle beam~\cite{Roth:JAS124:17}, while the density of the helium gas dropped
quickly~\cite{Horke:JAP121:123106}. Particles were detected 10~mm after the exit of the cell by
particle-localization microscopy based on optical light scattering~\cite{Awel:OptExp24:6507}. The
use of a light sheet to illuminate particles allowed a large-area illumination and hence direct
measurement of the entire transverse profile of the particle beam~\cite{Worbs:OptExp27:36580}. The
generated particle size distribution was monitored using a commercial differential mobility analyzer
(TSI 3786) and condensation particle counter (TSI 3081).

\subsection{Simulation details}
The experiments were complemented by quantitative simulations of nanoparticles traveling through the
apparatus. To model the gas-particle interactions within the buffer-gas cell, we developed a
numerical simulation framework capable of calculating the buffer-gas flow field, trajectories of
particles in the flow field, and the resulting particle temperatures. The velocities and pressures
of the helium flow-field were obtained by solving the Navier-Stokes equation at 4~K using a finite
element solver~\cite{Comsol:Multiphysics} for different mass flow conditions. Then, using a
homebuilt simulation framework, particle trajectories were calculated within the evaluated
steady-state flow field according to Stoke’s law. A temperature dependent particle-slip-correction
factor is required to calculate the drag forces~\cite{Willeke:JAS7:381}. As no such correction
factor was reported for cryogenic temperatures, we used the known values for air in the range
200--1000~K~\cite{Willeke:JAS7:381} scaled up by a factor of $4$ to give consistent results with our
experiment at cryogenic temperature. Due to the low nanoparticle densities, we assumed no effect of
the particles on the flow-field and no particle-particle interactions. Numerical integration is
performed using the Dormand \& Prince Runge-Kutta method \texttt{dopri5} as provided in
\texttt{scipy.integrate.ode}. The flowfield data are linearly interpolated using
\texttt{scipy.interpolate.RegularGridInterpolator}~\cite{Virtanen:NatMeth17:261}. The particles'
phase-space distribution at the inlet of the buffer-gas cell was assumed to be Gaussian, with mean
values and standard deviations obtained from simulating particle trajectories in the transport tube
and capillary using an cylindrically symmetric model for that part of the setup. Simulations through
the buffer-gas cell were performed using both, a two-dimensional (2D) description assuming
cylindrical symmetry and the three-dimensional (3D) exact experimental geometry. The latter was
deemed necessary because of small deviations of the apparatus from cylindrical symmetry due to the
precooled-helium inlet, see \autoref{fig:schematic}. At high helium flows this led to a noticeable
asymmetry in the produced particle distribution, which was well-reproduced by the 3D simulations,
\emph{vide infra}. Initial phase-space distributions of particles at the entrance of the buffer-gas
cell were taken from equivalent simulations of the transport system~\cite{Roth:JAS124:17}. The final
phase space distribution of the particle beam was collected at a detector placed 10 mm behind the
buffer-gas cell outlet.

Nanoparticle temperatures were evaluated by two independent approaches. A collision-based model was
used to calculate the temperature drop per helium-particle collision, ensuring conservation of
energy and momentum~\cite{Hutzler:CR112:4803}. This yields the particles translational temperature,
but does not take into account the thermal properties or the internal heat capacity of particles. In
the second approach the heat transfer from the nanoparticle into the buffer gas was estimated by
calculating the Nusselt number for forced convection of flow past a single
sphere~\cite{Whitaker:AIChE18:361}. The cooling rate taking into account the heat capacity of the
particles was then estimated according to Newton's law of cooling:
\begin{equation}
T(t)=T_\text{He}+(T(0)-T_\text{He})e^{-{hA}/{C}}
\end{equation}
with the temperature $T(t)$ of the particle at time $t$, $T_\text{helium}=4$~K, the initial
temperature of polystyrene $T(0)=298$~K, the surface area $A$ of the nanoparticle, the total heat
capacity $C$, which is the specific heat capacity $C_p$ multiplied by the particle mass, and the
heat transfer coefficient $h$. The latter was obtained by calculating the Nusselt number \Nuss for a
flow past a sphere using the Whitaker formula~\cite{Whitaker:AIChE18:361}:
\begin{equation}
\Nuss = 2 + \left(0.4\Reyn^{1/2}+0.06\Reyn^{2/3}\right)\Pran^{0.4}\left(\frac{\mu _{b}}{\mu_{0}}\right)^{\!\!1/4}
\label{eqn:good-name}
\end{equation}
with the Reynolds number \Reyn, the Prandtl number \Pran, the fluid viscosity $\mu_\text{b}$
evaluated at the bulk temperature $T_\text{He}=4$~K, and the fluid viscosity $\mu_0$ evaluated at
the initial surface temperature $T(0)=298$~K. As the mean free path of the helium gas is larger than
the nanoparticle diameter, a rarefied-gas correction was used~\cite{Kavanau:TransASME77:617}:
\begin{equation}
\Nuss = \frac{\Nuss_{0}}{1+3.42\frac{\Mach}{\Reyn\Pran}\Nuss_{0}}
\end{equation}
with the Nusselt number in the continuum regime $\Nuss_{0}$ and the Mach number \Mach. The heat
transfer coefficient $h$ was calculated as $h=k\Nuss/D$ with the diameter of the nanoparticle $D$
and the thermal conductivity of helium $k$.

\section{Results and discussion}
\label{sec:results}
\begin{figure}
   \includegraphics[width=\linewidth]{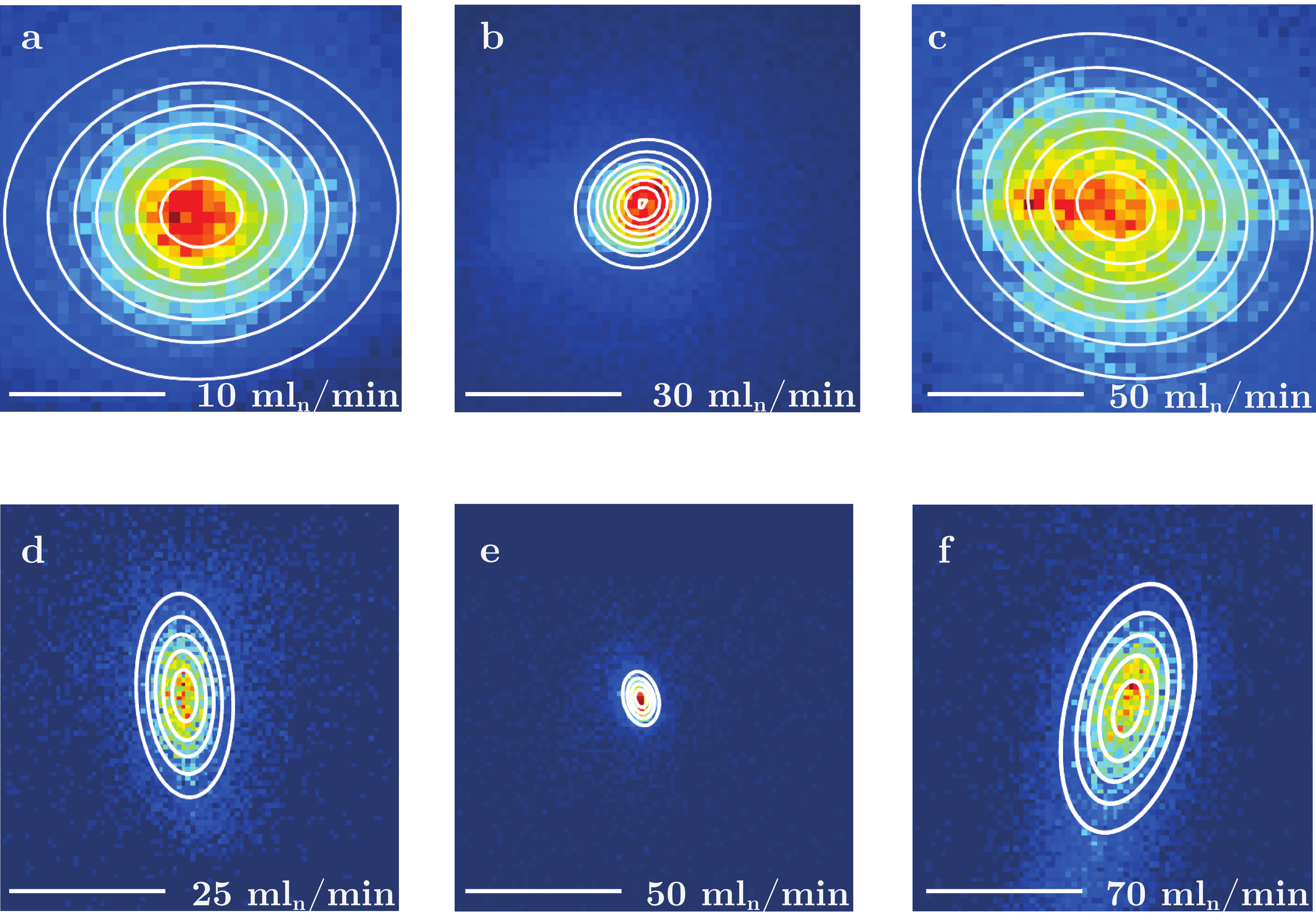}
   \caption{\textbf{Experimental particle beam profiles of polystyrene spheres.} Profiles of the
      particle beam emerging from the buffer-gas cell for different helium flow rates at the
      position of the light sheet \textbf{(a--c)} for 220~nm and \textbf{(d--f)} for 490~nm
      polystyrene spheres. The scalebars in the left bottom of the figure represent 500~\um, the
      individual helium gas flows are specified at the bottom right of every panel, and the color
      coding represent increasing particle flux from blue to red. Contour lines (white) represent 2D
      Gaussian fits; see text for details.}
   \label{fig:profiles}
\end{figure}
\begin{figure}
   \includegraphics[width=\linewidth]{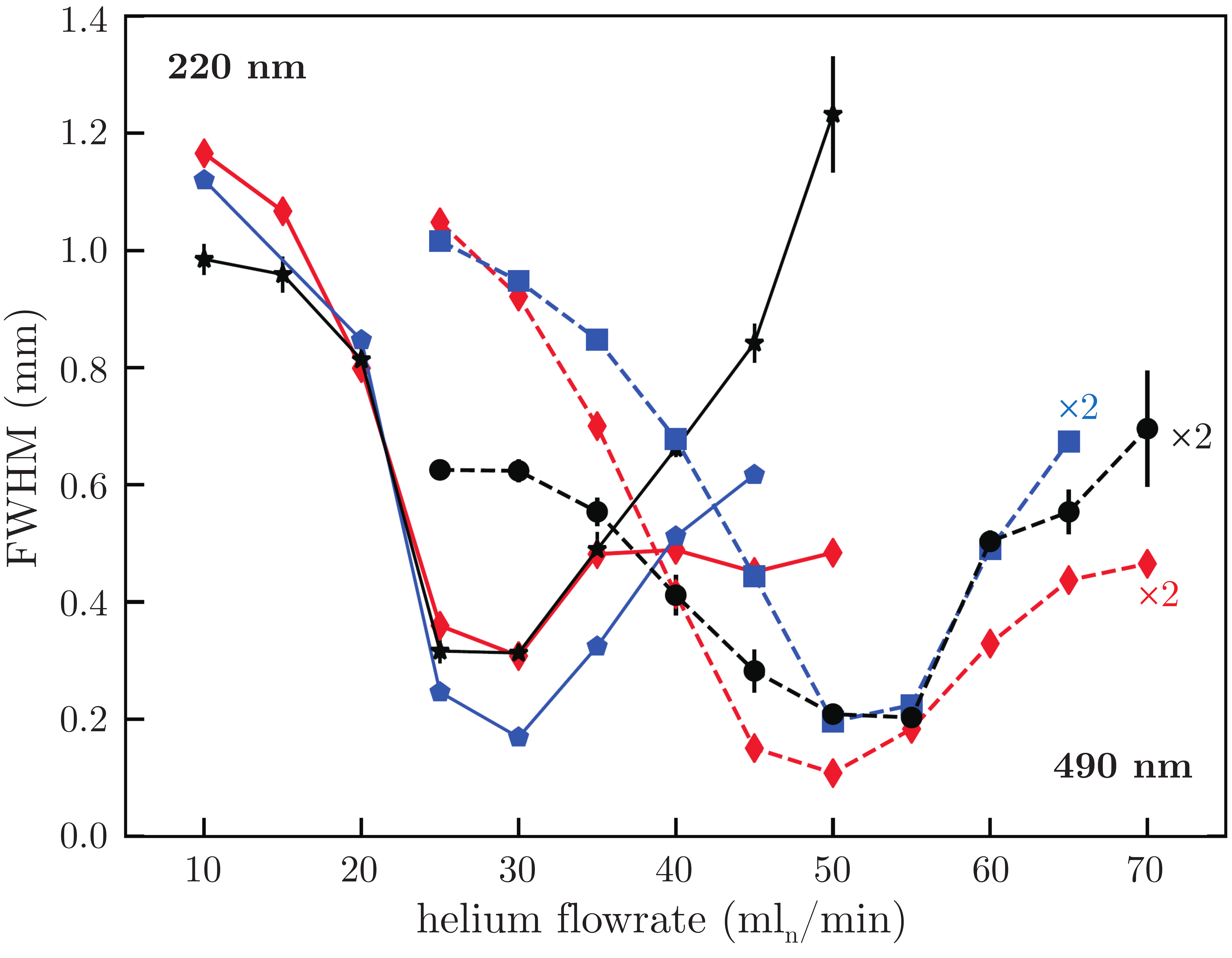}
   \caption{\textbf{Focusing behavior of polystyrene spheres.} Measured particle beam widths for
      490~nm (dashed lines) and 220~nm (solid lines) polystyrene spheres as a function of helium
      flow rate. Black lines represent the experimental data, while red lines are from
      two-dimensional-axisymmetric and blue lines from three-dimensional simulations, as discussed
      in the text. For the 490~nm datasets the width are scaled by a factor two to improve
      visibility of the variation. }
   \label{fig:focusing_curve}
\end{figure}

Spatial profiles of shock-frozen particles in the detection region are shown in
\autoref{fig:profiles} for 220~nm and 490~nm PS for different helium flow rates. The strong
variations of the particle beams for different flow conditions clearly indicate a strong
interaction, \ie, many collisions, with the helium gas. For the experimental detector position 10~mm
behind the cell outlet, the most collimated particle beam was observed at helium flow-rates of 30
and 50~\sccm for 220~nm and 490~nm PS, respectively. From \autoref{fig:profiles} it is evident that
the particle distributions were not spherically symmetric, but elliptical. We attribute this to an
asymmetric helium flow-field, caused by the location of our helium inlet at the top of the
buffer-gas cell inlet. Despite careful cell design, including a first gas inlet chamber for
providing a quasi-axisymmetric flow into the main cell~\cite{Singh:PRA97:032704}, at large flow
rates significant asymmetries existed in the gas flow, see \sifref{1}. We quantified the size of the
particle beams using a two-dimensional (2D) Gaussian, indicated by the contour lines in
\autoref{fig:profiles}. The measured dependence of the particle beam size on the helium flow is
shown in \autoref{fig:focusing_curve} (black curves). Here, we used the mean of the full width at
half maximum~(FWHM) of the minor and major axes of the 2D Gaussian to quantify the produced beam
size. Individual plots for the major and minor axis for both particle sizes are shown in \sifref{2}.
For both PS sizes an increase in helium flow led to a gradual decrease in particle beam size until
it reaches a minimum, \ie, a spatial focus, at the detector. Further increasing the helium flow
focused the particle beam further, moving the focus before the detector, which resulted in an again
larger beam size at the detector, as evident from simulated particle beam diameters at different
distances from the buffer-gas-cell outlet and for different flow conditions, see \sifrefs{3--S5}. We
simulated the measured focusing curves using both, 2D-axisymmetric and three-dimensional (3D)
asymmetric, flow-condition models, \emph{vide supra}. Comparisons between measured and simulated
beam widths for 220~nm and 490~nm PS are shown in \autoref{fig:focusing_curve}.
All simulations are in very good agreement with the experimental data. This also validated our
simulation framework, which thus provides further insight into the fluid-dynamic focusing process.
For instance, for 490~nm and 220~nm PS particles and a helium flow of 50~\sccm the simulations
yielded particle speeds in the laser-detection region of 16~m/s and 22~m/s, respectively. The
simulations also provided the phase-space distributions of the particle beams at different
coordinates within the buffer-gas cell, which for three different flow rates are visualized in
\sifrefs{4 and S5}. These distributions clearly illustrate the focusing effect, but also the
asymmetry present in the helium flow-field for large flows. While the obtuse angles of the
buffer-gas cell significantly reduce the formation of turbulences~\cite{Singh:PRA97:032704} the
asymmetry of the flow-field, with some indications of remaining turbulences, led to a significant
variation of the particles transverse velocities, especially at large helium flows. It is also
evident from the simulated particle beam diameters at different distances from the buffer-gas-cell
outlet, \sifref{3}, that at a very low helium flow of 25~\sccm not much focusing occurred and the
particle beam was collimated, in contrast to the typical convergence-divergence behavior at higher
helium flows, which resembles typical aerodynamic lens systems~\cite{Liu:AST22:293, Roth:JAS124:17}.
At sufficiently high flow rates, the thermalized particles in the buffer-gas cell followed the
flow-field and, when traveling through the small orifice, sped up. The large momentum of the
particles led to a more ballistic behavior when leaving the buffer-gas cell and thus a significantly
lower divergence of the particle beam than of the gas flow. The exact focusing properties of the
nanoparticle beam depended on the particles momentum and thus its fluid-dynamic properties and mass,
as well as the flow-field~\cite{Garrick:JAS37:555, Liu:AST22:293}. Generally, heavier particles
require larger gas flows for focusing. The particle transmission also increased with increasing
helium flow inside the cell, see \sifref{6}. For 220~nm particles, the maximum transmission is
achieved for a helium flow of 30--35~\sccm at 4~K, with a tenfold increase in transmission compared
to the lowest flow rate of 5~\sccm. This is attributed to stronger fluid-dynamic forces due to the
pressure increase, which efficiently guided the nanoparticles through the buffer gas cell and
minimized losses due to collisions with the walls~\cite{Hutzler:CR112:4803}. Flow conditions for
maximum transmission also coincide well with maximum focusing, yielding a seventy times higher flux
at the detector for 30~\sccm than for 5~\sccm, see \sifref{6}. With advanced fluid-dynamic focusing
outlets~\cite{Roth:JAS124:17}, beam focusing and particle flux can be improved even further.
Moreover, the effect of Brownian motion will be significantly reduced by the 4~K translational
temperature compared to previous room-temperature approaches. This is especially important for small
particles and thus will strongly improve their focusing and thus the densities in single-particle
imaging experiments.
\begin{table}
   \centering
   \begin{tabular}{lccccc}
     \hline\hline
     & 200 K & 133 K & 77 K & 10 K & Cooling rate \\
     & (\us) & (\us) & (\us) & (\us) & (K/s) \\
     \hline
     500 nm   & 613  & 1409 & 2467 & 12000 & 1.8$\times 10^5$ \\
     200 nm   & 224 & 476 & 821 & 3007 & 4.9$\times 10^5$\\
     50 nm   & 55 & 110 & 185 & 539 & 2.2$\times 10^6$ \\
     10 nm   & 12 & 23 & 37 & 103 & 1.1$\times 10^7$ \\
     Lysozyme & 6 & 10 & 16 & 40 & 2.6$\times 10^7$ \\
     \hline\hline
   \end{tabular}
   \caption{\textbf{Cooling rate in the buffer-gas cell for different particle sizes.} Calculated
      cooling rates at a fixed flow rate of 70~\sccm and the corresponding cooling times for
      reaching relevant temperatures, such as the protein glass-transition (200~K), water glass
      transition (133~K), and liquid nitrogen (77~K) temperatures. The cooling rates have an
      estimated error of 10~\%, propagated from the 10~\% error in the Nusselt
      number~[\onlinecite{Kavanau:TransASME77:617}].}
   \label{tab:temperature}
\end{table}
Our precise flow-field and particle-trajectory simulations allowed us to assess the temperature and
cooling rate of particles traveling through the cold buffer-gas cell. The number of collisions with
helium required for full thermalization depended on the thermal properties of the particle as well
as its size and velocity relative to the gas. In \autoref{tab:temperature}, we provide simulated
cooling times to several temperatures and corresponding initial cooling rates, for PS of 10--500~nm
diameter as well as for the prototypical protein lysozyme~\cite{Neutze:Nature406:752,
   Wiedorn:NatComm9:4025}. These were calculated assuming forced convection and Newton's law of
cooling and took into account the particles initial internal energy at room temperature. Full
cooling curves, \ie, the modeled temperature drop as a function of time as the particle traveled
through the buffer-gas cell and the instantaneous cooling rates are shown in \sifref{7} and S8,
along with results for a simpler momentum-transfer-based cooling model. These simulations show that
for particles smaller than $\ordsim50$~nm cooling rates on the order of $10^6$--$10^7$~K/s can be
achieved. This significantly exceeds the cooling rates for the plunge-freezing approach commonly
used in CEM~\cite{Dubochet:QRB21:129, Tivol:MicroscAna14:375}. Furthermore, the simulations show
that the cooling rate strongly depends on the initial position of the warm nanoparticle in the cold
cell, \ie, on the local helium density, and on the particles' velocity distribution. This provides
the way forward toward even faster cooling: Moving the position of the heated inlet capillary into
the buffer-gas cell will put the warm particles immediately into regions of higher-helium density.
Decoupling the initial-cooling cell from the fluid-dynamic focusing, \eg, in double-cell
configurations~\cite{Hutzler:CR112:4803}, would allow orders of magnitude higher densities of cold
helium at the inlet, providing correspondingly faster cooling. This two-cell setup will also enable
further control of the fluid-dynamic focusing at the outlet~\cite{Roth:JAS124:17}, enabling strongly
increased particle densities in the focus. Besides higher densities and better shock-freezing of
biological samples, such improvements and corresponding variability in the experimental parameters
would also enable studies of possible effects of the freezing rate on the structure of biological
macromolecules.

\begin{figure}
   \includegraphics[width=\linewidth]{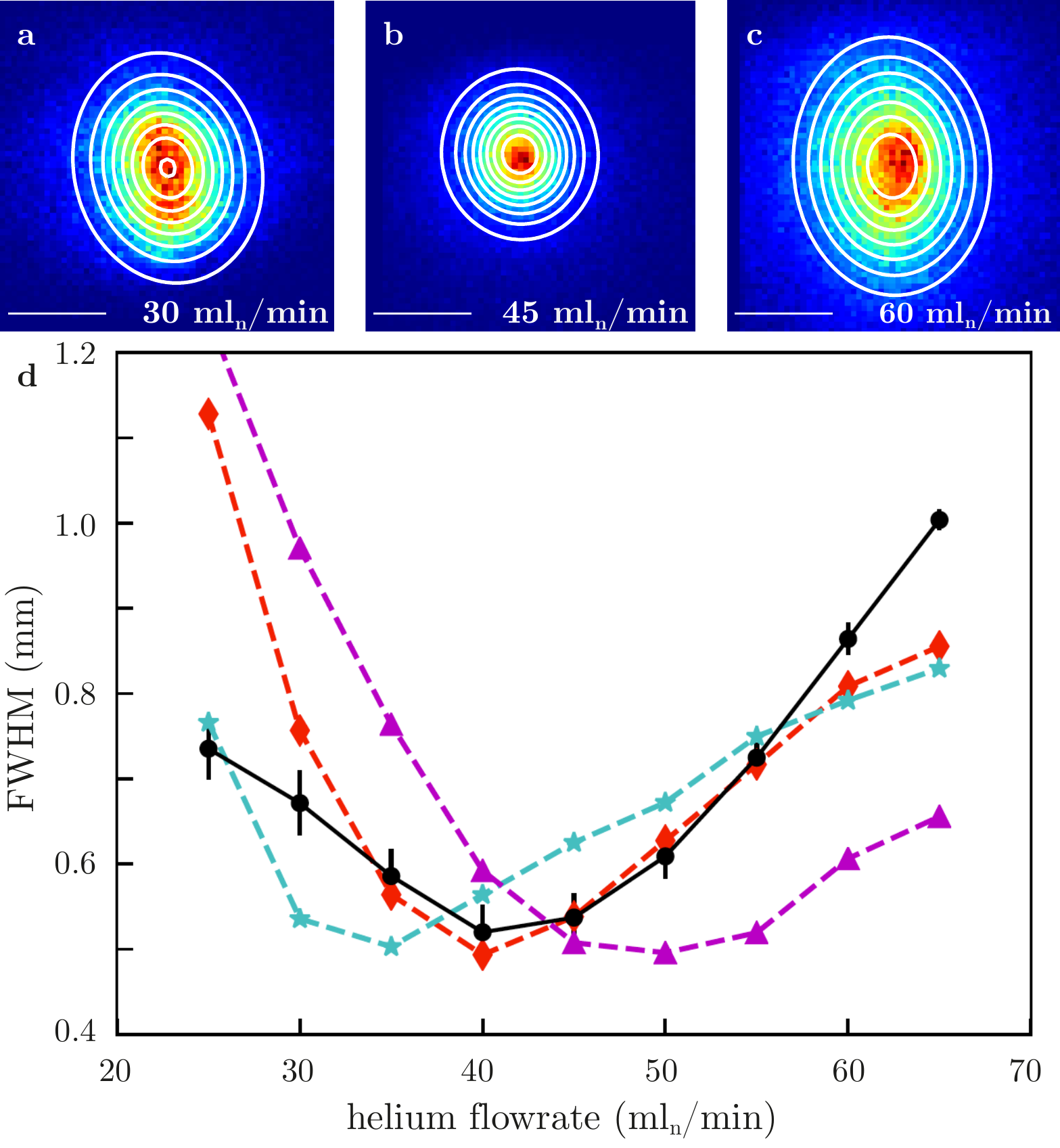}
   \caption{\textbf{Shock-freezing and focusing Granulovirus occlusion bodies.} \textbf{(a,~b,~c)}
      Experimental particle-beam profiles at the detector position for three different helium flow
      rates of 30, 45, and 60~\sccm, respectively. Scale bars and color codes are same as
      \autoref{fig:profiles}. \textbf{(d)} Experimental (black) and simulated (cyan, red and purple)
      particle beam widths as a function of helium flow rate. Simulations are shown for three
      different hydrodynamic diameters: 280~nm (cyan), 320~nm (red) and 360~nm (purple), confirming
      the expected effective hydrodynamic diameter of 320~nm for CpGV.}
   \label{fig:GV}
\end{figure}
To study the applicability of our approach to non-spherical biological nanoparticles, we created a
cryogenically cooled beam of CpGV. CpGV is readily available commercially as a insecticide
alternative to control codling moth populations~\cite{Gebhardt:PNAS111:15711}. The resulting
focusing curves following injection of CpGV into the buffer-gas cell are shown in \autoref{fig:GV}.
As previously observed for PS, the size of the produced particle beam at the detector showed a
strong dependence on the helium flow rate, with the narrowest profile and highest density observed
for helium flow rates of 40--45~\sccm. Our current simulation approach inherently assumes spherical
particles. Therefore, to model the CpGV data we simulated spherical particles of various sizes with
the known density of CpGV of 1160 $\text{kg}$/$\text{m}^3$. Simulations for diameters of 280~nm,
320~nm and 360~nm are shown in \autoref{fig:GV} and the best fit was observed for a particle size of
320~nm as an effective fluid-dynamic diameter for CpGV. This value is in good agreement with the
$325$~nm obtained as the geometric mean of CpGV's 3D diameters. This indicates that particles inside
the buffer-gas cell are freely rotating and no significant flow-alignment effects occur under the
experimental conditions of our study. Furthermore, these simulations confirm that also biological
particles leave the buffer-gas cell thermalized with the cold helium gas.

\section{Conclusion and outlook}
\label{sec:conclusion}
We have demonstrated a cryogenic nanoparticle source capable of producing tightly focused beams of
shock-frozen aerosolized nanoparticles and its quantitative description. Using a helium buffer-gas
cell, isolated room-temperature particles are rapidly cooled, typically reaching liquid nitrogen
temperatures within hundreds of microseconds, and quickly thermalizing with the buffer gas at 4~K.
The current outlet of the cell acts as a simple fluid-dynamic lens, efficiently extracting particles
and forming a focused beam. These beams were characterized through particle-localization microscopy.
The cooling and focusing properties can be tuned by varying the helium flow-rate and its
temperature. A novel numerical simulation infrastructure was set up to provide quantitative
simulations of particle trajectories and phase-space distributions, which are in very good agreement
with the measurements. These simulations then enabled the extraction of cooling rates and particle
temperatures, highlighting the very fast shock-freezing of nanoparticles. Last, but not least, we
demonstrated the applicability to non-spherical biological nanoparticles by producing beams of
shock-frozen granulovirus particles. Further improvements of the setup will provide orders of
magnitude faster cooling rates of the particles as well as better focusing of the emerging beams:
The initial cooling can be improved by placing the particle inlet into the buffer gas cell and
through two-cell approaches from small-molecule buffer-gas cooling~\cite{Hutzler:CR112:4803}. The
latter will also allow for advanced fluid-dynamic focusing outlets~\cite{Roth:JAS124:17} resulting
in strongly increased particle densities in the focus. The demonstrated high-flux beams of
shock-frozen nanoparticles will be beneficial to a wide range of experiments in structural biology,
nanoscience, and physics, including high-resolution single particle x-ray and electron
diffractive-imaging. In particular, our approach, together with control and selection, will overcome
the sample variability problem typically encountered in single-particle coherent x-ray diffraction
measurements, where millions of particles are needed to create a 3D
structure~\cite{Bergh:QRB41:181}. Furthermore, the beams of cold isolated particles open up a large
tool-box of control methods, originally developed for cold small gas-phase
molecules~\cite{Stapelfeldt:RMP75:543, Chang:IRPC34:557}, to these large nanoscale systems. These
include the separation of structural isomers or major folding structures~\cite{Chang:IRPC34:557,
   Trippel:RSI89:096110, Teschmit:ACIE57:13775} or molecular alignment approaches that fix molecules
in the laboratory frame using optical fields~\cite{Stapelfeldt:RMP75:543, Spence:PRL92:198102,
   Trippel:JCP148:101103, Amin:JPCL10:2938, Karamatskos:NatComm10:3364}. Such control would enable
the experimental averaging of imaging data over many identical molecules/particles. Furthermore, it
provides the prerequisites for future time-resolved studies of ultrafast biochemical dynamics, which
require well-defined starting states to controllably and reliably trigger specific dynamic processes
of interest. Additionally, the ability to control the particles final temperature and cooling rate
will allow the exploration of the ground-state potential energy landscape and answer important
outstanding questions regarding the preservation of native-like conditions upon rapid-freezing. It
furthermore enables the direct study of important temperature and size dependent phenomena in
artificial nanoparticles, such as extremely large magnetoresistance~\cite{Ali:Nature514:205} or
light-induced superconductivity~\cite{Mitrano:Nature530:461}. Furthermore, it could propel
matter-wave interference to new limits~\cite{Eibenberger:PCCP15:14696}.

Our approach enables imaging experiments that bring the benefits of CEM, well-controlled and static
sample particles, to single-particle imaging where they can be imaged \emph{in vacuo} without
support structures and with ultrafast time-resolution. In turn, combining the very fast cooling
enabled by our approach with soft-landing techniques could bring strong and crucial progress to the
sample delivery in CEM experiments.

\section{Supplementary Materials}
\noindent See the supplementary material for
\begin{description}
\item[Fig.~S1.] Helium flowfields for different flow conditions in the buffer-gas cell.
\item[Fig.~S2.] Measured dependence of the particle beam size on the helium flow rates for
   polystyrene spheres.
\item[Fig.~S3.] Simulated focusing behavior at different helium flow rates.
\item[Fig.~S4.] Phase space distributions of a beam of 490~nm PS particles 5~mm before the
   buffer-gas cell outlet.
\item[Fig.~S5.] Phase space distributions of a beam of 490~nm PS particles 5~mm after the buffer-gas
   cell outlet.
\item[Fig.~S6.] Experimentally obtained transmission and average flux of particles.
\item[Fig.~S7.] Simulated thermalization times of nanoparticles.
\item[Fig.~S8.] Rate of cooling as a function of temperature.
\end{description}

\section*{Acknowledgments}
The expert technical support by Tim Ossenbrüggen, Nicolai Pohlmann, and Karol Długołeęcki is
gratefully acknowledged. We also thank Dominik Oberthür for the supply of purified CpGV particles.

This work has been supported by the European Research Council under the European Union's Seventh
Framework Programme (FP7/2007-2013) through the Consolidator Grant COMOTION (ERC-614507-Küpper), by
the Helmholtz Gemeinschaft through the ``Impuls- und Vernetzungsfond'', and by the Clusters of
Excellence at Universität Hamburg, the ``Center for Ultrafast Imaging'' (CUI, EXC~1074,
ID~194651731) and ``Advanced Imaging of Matter'' (AIM, EXC~2056, ID~390715994) of the Deutsche
Forschungsgemeinschaft (DFG).

\paragraph{Author contributions}
The project was conceived by JK and coordinated by DAH and JK. The experiment was designed by AS,
NR, DAH, and JK; set up by AS, AE, and LW; and performed by AS and AE. The data analysis was
performed by AS and DAH and numerical simulations were performed by MA, NR, and AS; the results from
theory and experiment were analyzed by AS, MA, DAH, and JK and discussed with all authors. The
manuscript was prepared by AS, DAH, and JK and discussed by all authors.

\bibliography{string,cmi}
\end{document}